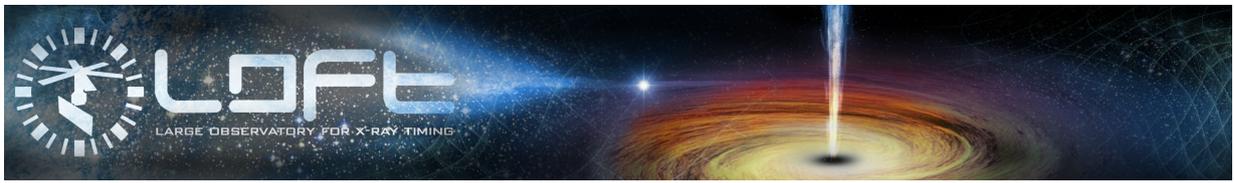

# Dissecting accretion and outflows in accreting white dwarf binaries

White Paper in Support of the Mission Concept of the
Large Observatory for X-ray Timing

Authors

D. de Martino[1], G. Sala[2], S. Balman[3], F. Bernardini[1,4], A. Bianchini[5], M. Bode[6],
J.-M. Bonnet-Bidaud[7], M. Falanga[8], J. Greiner[9], P. Groot[10], M. Hernanz[11], G. Israel[12],
J. Jose[2], C. Motch[13], M. Mouchet[14], A.J. Norton[15], A. Nucita[16], M. Orio[17], J. Osborne[18],
G. Ramsay[19], P. Rodriguez-Gil[20], S. Scaringi[9,21], A. Schwope[22], I. Traulsen[22], F. Tamburini[5]

[1] INAF – Osservatorio Astronomico di Capodimonte, Naples, Italy
[2] Univ. Politecnica de Catalunya and IEEC, Barcelona, Spain
[3] Middle East Technical University, Ankara, Turkey
[4] New York University Abu Dhabi, Abu Dhabi, United Arab Emirates
[5] Universitá di Padova, Padova, Italy
[6] Liverpool University, Liverpool, United Kingdom
[7] CEA Saclay, DSM/Irfu/Service d'Astrophysique, Gif-Sur-Yvette, France
[8] International Space Science Institute (ISSI), Bern, Switzerland
[9] Max Planck Institut für Extraterrestrische Physik, Garching, Germany
[10] Radboud University, Nijmegen, The Netherlands
[11] Institute of Space Sciences, ICE (CSIC-IEEC), Barcelona, Spain
[12] INAF – Osservatorio di Monteporzio, Rome, Italy
[13] Observatoire Astronomique de Strasbourg, Strasbourg, France
[14] LUTH, Observatoire de Meudon and Université Paris Diderot, Paris, France
[15] The Open University, Walton Hall, United Kingdom
[16] Universitá del Salento, Lecce, Italy
[17] INAF – Osservatorio Astronomico di Padova, Padova, Italy
[18] University of Leicester, Leicester, United Kingdom
[19] Armagh Observatory, Armagh, United Kingdom
[20] Instituto de Astrofisica de Canarias, La Laguna, Spain
[21] Institute of Astronomy, Leuven, Belgium
[22] Leibniz-Institut für Astrophysik Potsdam (AIP), Potsdam, Germany

**Preamble**

The Large Observatory for X-ray Timing, *LOFT*, is designed to perform fast X-ray timing and spectroscopy with uniquely large throughput (Feroci et al. 2014). *LOFT* focuses on two fundamental questions of ESA's Cosmic Vision Theme "Matter under extreme conditions": what is the equation of state of ultra-dense matter in neutron stars? Does matter orbiting close to the event horizon follow the predictions of general relativity? These goals are elaborated in the mission Yellow Book (http://sci.esa.int/loft/53447-loft-yellow-book/) describing the *LOFT* mission as proposed in M3, which closely resembles the *LOFT* mission now being proposed for M4.

The extensive assessment study of *LOFT* as ESA's M3 mission candidate demonstrates the high level of maturity and the technical feasibility of the mission, as well as the scientific importance of its unique core science goals. For this reason, the *LOFT* development has been continued, aiming at the new M4 launch opportunity, for which the M3 science goals have been confirmed. The unprecedentedly large effective area, large grasp, and spectroscopic capabilities of *LOFT*'s instruments make the mission capable of state-of-the-art science not only for its core science case, but also for many other open questions in astrophysics.

*LOFT*'s primary instrument is the Large Area Detector (LAD), a $8.5\,\mathrm{m}^2$ instrument operating in the 2–30 keV energy range, which will revolutionise studies of Galactic and extragalactic X-ray sources down to their fundamental time scales. The mission also features a Wide Field Monitor (WFM), which in the 2–50 keV range simultaneously observes more than a third of the sky at any time, detecting objects down to mCrab fluxes and providing data with excellent timing and spectral resolution. Additionally, the mission is equipped with an on-board alert system for the detection and rapid broadcasting to the ground of celestial bright and fast outbursts of X-rays (particularly, Gamma-ray Bursts).

This paper is one of twelve White Papers that illustrate the unique potential of *LOFT* as an X-ray observatory in a variety of astrophysical fields in addition to the core science.

# 1 Summary

Accreting white dwarf binaries, cataclysmic variables, symbiotics, double-degenerates and novae, represent ideal astrophysical environments to study accretion and outflow processes in a wide variety of plasma conditions. Among the white dwarf accretors one finds the possible progenitors of Supernovae type Ia. They are sources of X-ray emission whose properties strongly depend on fundamental parameters such as the accretion rate, the mass and the magnetic field of the accreting white dwarf. Thus, understanding the processes of X-ray emission and the evolution of close accreting white dwarf binaries is of utmost importance in a wider context. The X-rays powerfully diagnose the conditions shaping the accretion flow close to the white dwarf as well as the mechanisms of mass ejection and energetics. In recent years accreting white dwarf binaries have surprisingly been found to be hard X-ray emitters with a huge variety of spectral and variability characteristics. Understanding the underlying physical processes requires high sensitivity and timing capabilities as offered by the *LOFT* mission. The three main questions to be addressed are:

- *How does matter accrete onto white dwarfs? LOFT* will address this by studying selected samples of ten magnetic and non-magnetic systems each to detect for the first time low-amplitude fast aperiodic and periodic variabilities in the hard X-rays and performing phase-resolved spectroscopy.

- *How does mass ejection work in nova explosions? LOFT* will investigate this by monitoring ~three novae per year to detect the onset of hard X-ray emission and its spectral variability along outburst as well as to characterise for the first time fast variability in the hard X-ray regime.

- *What causes dwarf novae outburst diversity and what are the conditions for disc-jet launching? LOFT* will address this by following ~three dwarf novae per year through time-resolved spectroscopy and fast timing in the hard X-rays with unprecedented details.

Thus *LOFT* will be crucial to measure the spectral and temporal properties of the poorly known hard X-ray tails in these systems. When *LOFT* is expected to be operational, wide field area surveys will have provided statistically significant samples to allow detailed investigation in coordination with the large facilities foreseen in the post-2020 timeframe.

# 2 Introduction

Accreting white dwarf (WD) binaries represent the most common end products of close binary evolution. They encompass various classes of systems: the cataclysmic variables (CVs) in which the donor (secondary) star is a late type main sequence star in a close orbit ($P_{\rm orb} \sim 1.4\text{–}14\,{\rm hr}$), the AM CVn stars where the donor is a (semi)degenerate He-rich star in an ultracompact binary ($P_{\rm orb} \sim 10\text{–}65\,{\rm min}$) and the symbiotics where the donor star is a late type giant or Mira star in wide ($P_{\rm orb} \sim$ years) orbit. Mass transfer occurs either through Roche-Lobe overflow in the CVs and AM CVn stars or via stellar wind in the symbiotics. These binaries display a wide range of timescales of variabilities (from sec–mins up to months–years) that make them an ideal laboratory to study accretion/ejection processes in non-relativistic regimes. Disc instabilities, disc-jet connection and magnetic accretion, that are common processes occurring in a wide variety of astrophysical objects from pre-main sequence stars to BH binaries, make accreting WDs a challenging link to unveil universal connections. Furthermore, thermonuclear runaways occurring during nova explosions have the potential to unveil the accretion/ignition mechanisms, the chemical enrichment of the interstellar medium as well as the effects of accretion onto the WD that may drive its mass to grow over the Chandrasekhar limit and explode as SN Ia, a key aspect for our understanding of the progenitors. In addition, population studies of galactic X-ray sources have recently recognized a crucial role of accreting WD binaries. Of particular importance is the origin of the Galactic Ridge X-ray Emission (GRXE), one of the great mysteries in X-rays, thought to be composed of a non-negligible population of CVs (Revnivtsev et al. 2009; Revnivtsev et al. 2011; Yuasa et al. 2012; Hong 2012; Warwick 2014).

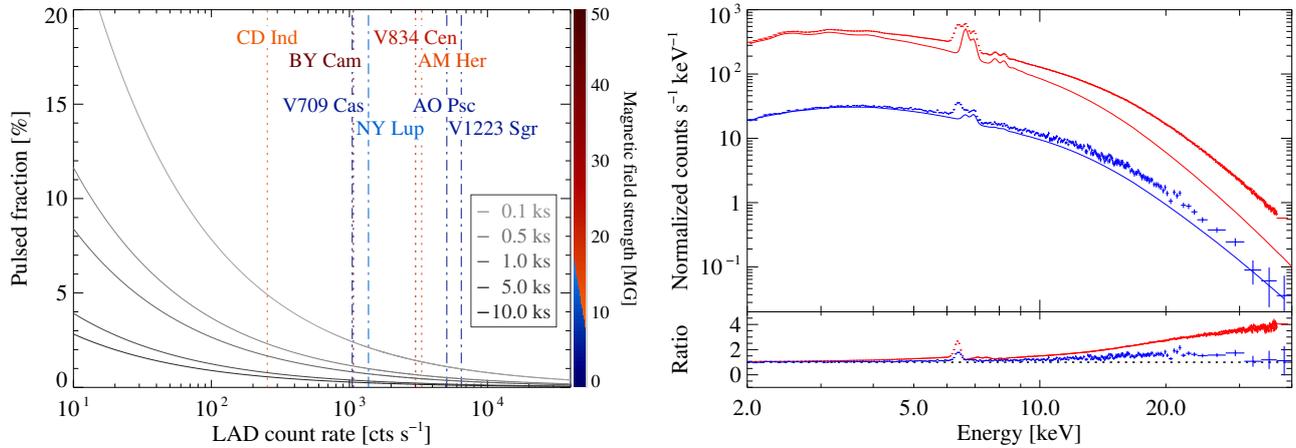

Figure 1: *Left:* Feasibility of detecting fast ~1 Hz QPOs in LAD light curve segments of different length (from top to bottom: 0.1–10 ks). The grey lines show the minimum amplitudes vs LAD count rates for which periodicities can be detected at a $4\sigma$ confidence level, based on $Z_1^2$ statistics (Buccheri et al. 1983). Vertical lines show the expected LAD count rates for a sample of X-ray hard mCVs during bright states, derived from power-law fits to Swift/BAT spectra from Baumgartner et al. (2013). The sample covers a wide range of magnetic field strengths from 11 MG (CD Ind) to 40 MG (BY Cam) for the Polars and $B \lesssim 10$ MG for the Intermediate Polars. *Right*: Simulated 10 ks LAD spectra of a faint mCV (blue; RX J2133+51, after model B2 of Anzolin et al. 2009) and a bright polar (AM Her) (red) with $F_{3-20\,\text{keV}} = 3.5 \times 10^{-11}$ and $2.1 \times 10^{-10}$ erg cm$^{-2}$ s$^{-1}$, respectively. Both composite spectra comprise absorption, a multi-temperature post-shock region, a Compton reflection continuum and a Gaussian Fe K$\alpha$ fluorescent line. The residuals compared to models not accounting for reflection (lower panel) elucidate the opportunities opened by *LOFT* even for faint objects and the prospects for phase-resolved spectroscopy.

The X-ray domain is crucial to understand the physical condition of matter close to the compact star. Despite important progress achieved in recent years, there are many open questions that still deserve full exploitation, but hampered by the low quiescent luminosities ($L_X \sim 10^{29-33}$ erg s$^{-1}$) and the unpredictability of large-scale variations, that can reach luminosities of $\sim 10^{32} - 10^{34}$ erg s$^{-1}$ in dwarf novae (DNe) outbursts and up to $\sim 10^{38}$ erg s$^{-1}$ in Nova explosions. The unique combination of high temporal, moderate spectral capabilities and sensitivity of the
lad instrument, as well as the wide field coverage of the WFM instrument onboard *LOFT*, will be crucial to uncover the wide range of variabilities in the mostly unexplored hard X-ray domain.

The major scientific goals are here summarized for accreting WD binaries for which *LOFT* will provide breakthrough results. In Sect. 3 we will address how *LOFT* will allow to study X-ray oscillations and aperiodic variability to probe details of the accretion flow and its instabilities in CVs. In Sect. 4 we will discuss the importance of the evolution and onset of hard X-ray emission in novae. In Sect. 5 we will address *LOFT*'s capability to investigate the large diversity in the X-ray behaviours in DN outbursts and whether in these scaled-down systems of NS/BH binaries a disc-jet connection is also at work. Finally, in Sect. 6 the importance of such X-ray mission will be discussed in the context of foreseen facilities in other wavelength domains in the post-2020 time frame.

## 3 How does matter accrete onto WDs? Probing accretion through oscillations

### 3.1 Accretion in WDs

Accretion of matter onto WD can take place through a disc, a ring-like structure or an accretion stream depending on the WD magnetic field strength, the binary orbital period and mass accretion rate. The physical conditions of the material impinging onto the WD and emitting X-rays are still to be investigated. In non-magnetic CVs the WD magnetic field strength is low enough, $B_{WD} \lesssim 10^5$ G, to allow the formation of an accretion disc and X-rays are emitted at the disc Boundary Layer (BL). In highly magnetised CVs (mCVs), $B_{WD} \gtrsim 10^7$ G, material is directly accreted onto the WD poles flowing along the magnetic field lines. A partial accretion disc can be formed at intermediate field strengths. At the WD poles a stand-off shock is formed. The post-shock accretion column is thought to cool via bremstrahlung (hard X-rays), recombination processes and cyclotron radiation. The relative proportion of these cooling mechanisms strongly depends on the WD magnetic field (Fischer & Beuermann 2001).

Both the BL and the shock radiate in the X-rays a substantial fraction of gravitational energy and thus their structure and stability can be efficiently probed through spectral and temporal X-ray properties. Of particular importance is the WD mass that can be determined even in faint sources, thanks to the extended coverage and unique sensitivity of *LOFT* in the hard X-rays. This is crucial to understand the role of WD binaries as SNIa progenitors. Mostly unexplored are non-periodic variations that uncover a wide range of timescales (from a few sec to hundreds-thousands of sec). These are observed up to now mainly at optical wavelengths, in both non-magnetic and magnetic systems. Searches of X-ray counterparts were carried out in a few bright systems leading to partial, sometimes controversial results.

### 3.2 QPOs in magnetic cataclysmic variables

One of the most challenging aspects is the origin of 1–10 s quasi-periodic-oscillations (QPOs) in mCVs. Optical narrow QPOs with a few percent amplitude were detected in some mCVs of the polar type (Middleditch et al. 1997), but not in all of them. If these represent oscillations arising in the post-shock region they should be detected in the X-rays as well. Attempts to search the X-ray counterparts has however failed so far, providing only upper limits ($\lesssim$ 10–20%) (Beardmore & Osborne 1997; Imamura et al. 2000; Pandel & Cordova 2002; Ramsay et al. 2007; Bonnet-Bidaud et al. 2014), mainly due to the lack of adequate sensitivity in the hard X-rays. The origin of QPOs in mCVs is debated, but has great potential to allow an understanding of radiative shocks. Extended shocks are predicted to be thermally unstable giving rise to QPOs, in both optical and X-ray domains. QPO characteristics (frequency, amplitude and phase) are related to the cooling timescales of the emitting plasma (Wu 2000). Pure radiative shocks were however strongly questioned on the basis of stability studies. In particular, in the presence of strong magnetic fields ($\gtrsim 10^7$ G), cyclotron cooling has a significant role in damping oscillations. Alternative inhomogeneous accretion or noise-driven shock oscillation models were also invoked (Steiman-Cameron et al. 1994). An important constrain on the models can be provided by variations of amplitude and phase of the oscillations and their energy dependence. The lower amplitudes of the 0.3-1 Hz QPOs together with their strong dependence on the rotational phase made, up to now, the detection a challenge because it requires high sensitivity in the hard X-rays, where bremsstrahlung dominated shocks emit most. **The high sensitivity and energy coverage of the LAD will be then crucial to detect fast X-ray QPOs in mCVs down to amplitudes of 1%** (Fig. 1, left panel). To fully test the shock structure stability, a well defined sample of ~10 hard X-ray emitting mCVs covering a wide range of WD magnetic field strengths is needed. This should include moderate field Polars and the low-field Intermediate Polars to be observed along their rotational cycles (few tens of minutes to a few hours, depending on their magnetic field strength). The sample should also include the few known eclipsing systems to allow direct measures of the size and brightness distributions over the accretion spot thus uncovering the complex interplay between the magnetic field and matter in the accretion regions. In this way it will be possible to probe shock stability against a wide range of physical conditions of the post-shock region to be traced with phase-resolved spectroscopy along the spin cycle (Fig 1, right panel). Particularly important will be the ability to constrain the Compton reflection components, ubiquitous in these systems, through the fluorescence Fe K$\alpha$ line and continuum that peaks in the 10–30 keV range.

### 3.3 Rapid Oscillations in dwarf novae

Rapid flux oscillations (DNOs) of ∼3–40 s are observed in high mass accretion rate CVs – mostly during DN outbursts (see also Sect. 5) and mainly at optical wavelengths with a few detections in the EUV or soft X-rays (see review in Warner 2004). The optical DNOs are found to have high coherence and low amplitude (∼ $1 - 2\%$). They relate to the outburst evolution, appearing during the rise, persisting at maximum and disappearing during the decline of outbursts. Their period decreases during the rise to the maximum, while it increases during the outburst decay. The few systems monitored in the X-rays show that DNOs during outbursts occur in the soft X-rays. On the other hand, low-coherence and long period (mins) QPOs with amplitudes ∼ $10 - 40\%$ are mainly observed in the hard X-rays and generally appear during the hard X-ray turn-on at the end of optical outbursts (Wheatley et al. 1996). Their characteristic periods and amplitudes vary along the outburst, but these changes do not seem to be related to spectral variations (Wheatley et al. 2003).

The optical QPOs and DNOs (period ratio 15) correlate as the low and high-frequency QPOs of LMXBs strengthening the evidence of a general relation in all interacting binaries, irrespective of whether the primary is a BH, NS or WD (Warner et al. 2003). They are thought to be manifestations of disc accretion onto a very low-magnetic field compact object. The optical DNOs are believed to be due to magnetic coupling in an equatorial belt while the QPOs to reprocessing in the accretion disc or oscillation in the disc itself (Warner 2004). The hard X-ray QPOs cannot have the same origin and may reflect variations in the mass accretion rate at the inner BL. Their appearance is related to the transition from optically thick to optically thin of the BL and thus have a great potential to infer details of BL regimes. The study of fast oscillations in the X-rays is thus crucial but hampered so far due to the lack of X-ray coverage over a broad energy range and adequate monitoring of the different evolution of outbursts (see Sect. 5 for outburst diversity). *LOFT* will be able to detect the oscillations and monitor the frequency change along outburst even when sources will reach quiescent fluxes.

### 3.4 Aperiodic variability

Broad-band variability (flickering) is known to be an ubiquitous feature amongst all accreting compact binaries. In particular, both X-ray and optical aperiodic variabilities in CVs have been recognised for over a decade. A "fluctuating accretion disc model", where variations in the mass transfer rate generated throughout the disc couple, multiplicatively, to give rise to the observed flickering, was originally developed for X-ray binaries (XRBs) and Active Galactic Nuclei (AGN) (Arevalo et al. 2006; Ingram & van der Klis 2013). This model was successfully applied to high S/N optical lightcurves of the nova-like MV Lyr (Scaringi et al. 2012; Scaringi 2014). It was also applied to the X-ray light curves of a few non-magnetic CVs in quiescence (Balman & Revnivtsev 2012) allowing to infer disc truncation from the observed break in the X-ray power spectra. A possible influence of the weak magnetic fields in these systems was also claimed (Anzolin et al. 2010). The X-rays, especially the hard range, diagnose the inner disc regions and hence are crucial to establish the driving mechanism and whether discs are truly truncated in quiescence. This requires much higher S/N achieved so far, because these systems are rather faint in quiescence. High quality and high temporal resolution light curves will be obtained by *LOFT* allowing the first fair comparison of X-ray flickering in CVs with those of XRBs. *LOFT* thus holds the potential to unify aperiodic variability across all accreting compact sources, irrespective of the accretor type.

To study both X-ray QPOs and flickering in non magnetic CVs, a sample of ∼10 systems needs to be explored both in quiescence and outburst (see also Sect. 5). The sample should encompass a range of orbital periods, hence of mass transfer rates, to overcome selection effects and to define the statistical properties of non-periodic variabilities. Fig. 3, right panel, shows the great capability of *LOFT* to detect QPOs and the frequency break expected for a source at comparable flux levels of SS Cyg during the turn-on of hard X-rays at the end of the optical outburst.

## 4 How does mass ejection work in nova explosions?

Nova outbursts occur when the pressure at the base of the accreted envelope on the WD in a close binary system is high enough to initiate a hydrogen thermonuclear runaway that leads to the ejection of a part of the accreted envelope; the explosion does not disrupt the WD, so that the classical nova (CNe) phenomenon recurs every $10^3$–$10^5$ years. Recurrent novae (RNe) show outbursts recurring over a human life timescale. About half of them have long orbital periods with a red giant companion (symbiotic RNe, SyRN). RNe are thought to host a massive WD, close to the Chandrasekhar limit and their high mass transfer rate enable frequent outbursts. They are thus considered candidate SNIa progenitors, with the final fate of the WD depending on the balance between the mass accretion and ejection processes.

Novae are luminous X-ray sources (see statistics and reviews in Orio et al. 2001; Orio et al. 2012; Schwarz et al. 2011), with emission spanning from the very soft to the hard X-ray energy range. While the soft X-rays probe the continuous nuclear burning on the H-rich envelope on the WD surface, the hard X-rays are believed to trace mass outflows and the onset of resumed accretion (Mukai & Ishida 2001; Bode et al. 2006; Sokoloski et al. 2006; Nelson et al. 2012; Orio et al. 2014). The details of the mass outflow, the shaping and evolution of the ejected shell are still very poorly known. Recent efforts in the study of the ejecta morphology include modelling of the optical and X-ray emission line profiles and radio imaging (O'Brien et al. 2006; Orlando et al. 2009; Ribeiro et al. 2013; Chomiuk et al. 2014), but better quality X-ray data at an early post-outburst stage will be essential to progress.

Our lack of a detailed knowledge of the processes occurring in the expanding ejecta has been recently evident with the unprecedented detection by Fermi/LAT of very high energy (VHE) gamma rays at energies larger than 100 MeV in four novae, during the first 2–3 weeks after outburst (Ackermann et al. 2014). In fact, particle acceleration early after the explosion was predicted for the SyRN RS Oph 2006 outburst (Tatischeff et al. 2007). Such a mechanism produces gamma rays, which could not be deceted before the advent of Fermi in 2008. RS Oph showed copious X-ray flux in the 10–50 keV range from the beginning of the outburst and for about 3 weeks, with *RXTE* and *Swift*/BAT (Sokoloski et al. 2006; Bode et al. 2006). The fact that the first gamma-ray nova detected by Fermi/LAT was the SyRN V407 Cyg 2010 (Abdo et al. 2010) was initially interpreted as being a process related to the strong shocks of the expanding ejecta colliding with the dense red giant wind. However, the additional Fermi-LAT detections of three more novae without giant companions show that the acceleration of particles to high energies also occurs in the CNe ejecta, without invoking the dense red giant wind of the SyRN. The further detection of evolving radio emission from nova V959 Mon has recently been shown to be a new tool to pinpoint the location of the particle acceleration responsible for the gamma-rays detected by Fermi (Chomiuk et al. 2014). Thus, the monitoring of the early hard X-ray emission of novae in coordination with radio coverage has the unique potential to understand whether hard X-rays are truly related to the processes of particle acceleration.

The central supersoft X-ray source shows variability and periodicities on many different timescales, ranging from days to less than a minute. Periodicities of several hours can be attributed to orbital modulation, as in V5116 Sgr (Sala et al. 2008) while those with timescales of fractions of hours can be related to the WD spin. Variations at ∼19–37 min were found in some novae also in M31 (Leibowitz et al. 2006; Dobrotka & Ness 2010; Ness et al. 2011; Pietsch et al. 2011)). In contrast, a transient 35 s modulation found in the soft X-ray emission of RS Oph was suggested to be related to H-burning instabilities (Ness et al. 2007; Osborne et al. 2011). Such short period oscillations have now been found in several other novae. However, the lack of enough sensitivity has prevented high-resolution timing studies in the hard X-rays originating in the ejecta and, possibly, in the process of resuming accretion.

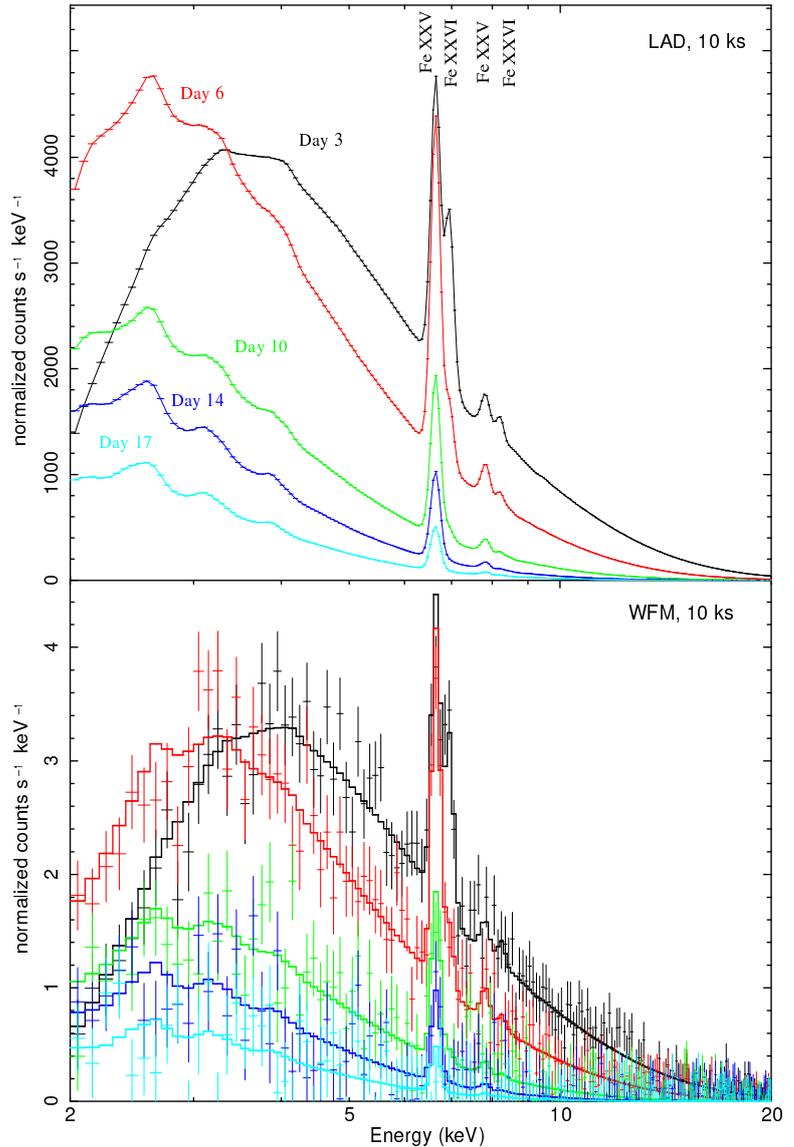

**Figure 2:** Simulations of the spectral evolution of the SyRN RS Oph over its outburst, both as seen by LAD (upper panel) and by the WFM (lower panel). The spectra include a thermal plasma model (*mekal*) with solar abundances, evolving from $kT = 8.2\,\mathrm{keV}$ and a flux of $3.2 \times 10^{-9}\,\mathrm{erg\,cm^{-2}\,s^{-1}}$ (0.5–20 keV) on day 3, down to $kT = 2.5\,\mathrm{keV}$ and a flux of $4.3\times10^{-10}\,\mathrm{erg\,cm^{-2}\,s^{-1}}$ (0.5–20 keV) on day 17 (after Sokoloski et al. 2006). The high X-ray flux during the first 17 days would trigger the WFM in full resolution mode and detailed spectral evolution will be followed by the LAD.

*LOFT* will be uniquely able to both monitor with the WFM the sky for bright nova outbursts, and follow them with LAD to characterise spectral variability as well as fast variabilities in the hard X-ray emission that would yield information on the processes occurring during mass ejection. The brightest and fastest RN, with flux of the same order of magnitude as that of RS Oph (located at 1.6 kpc, Hjellming et al. 1986), would trigger the WFM even at a distance of 5 kpc immediately after the outburst (Fig. 2, lower panel). Thus the WFM with its large field of view is an ideal instrument to discover them. For novae not bright enough to trigger the WFM, LAD observations would be triggered few days after optical discovery and analysis with some pre-established trigger criterion (based on distance, and/or fast evolution of the nova). Simulations show that the early hard X-ray emission (as for example detected in RS Oph with RXTE and Swift/BAT (Sokoloski et al. 2006; Bode et al. 2006)) will provide a high rate in the LAD (Fig. 2, upper panel) to allow fast timing studies for the continuum on timescales shorter than a second or a few mins for fluxes usually detected in hard X-rays for novae in outburst, as well as to infer spectral variations (e.g., the Fe line complex) down to tens of minutes. In this way it will be possible to probe in details the ejection process and the shocks in the expanding shells during the outburst.

Thanks to the Swift observations of novae in the last few years, we know that around half of the discovered Galactic novae are hard X-ray sources (in the *Swift*/XRT hard band up to 10 keV). This represents *a total of 2–3 novae per year*. The LAD's exceptional effective area would provide spectra with high statistics for all of them, making possible the first accurate study of a sample of ∼10 novae during the lifetime of *LOFT*. Furthermore, whether some novae remain hard X-ray emitters several years after eruption, possibly due to a magnetic nature of the WD, is an exciting issue (Hernanz & Sala 2002) that *LOFT* will be able to address.

## 5   What causes DN outburst diversity and what are the conditions for disc-jet launching?

DN outbursts ($L_X \sim 10^{32}$–$10^{34}$ erg s$^{-1}$) are believed to be due to disc instabilities. The knowledge of viscosity is crucial to understand angular momentum transport in accretion discs. Progress can be made by studying those phenomena that change on viscous timescales. The great advantage of DNe are their outburst timescales (hence viscous timescales) of only a few days or weeks, which are, thus, easy to access. DN outbursts are characterized by optical brightening by 2–5 mag lasting typically for a week and recurrence time of several tens of days. Much longer outburst recurrence times occur in low mass accretion rate systems. The outburst cycles are understood as due to disc instabilities (Lasota 2001). However very little is known about the inner disc regions and how they change from outburst to quiescence. These are best traced in the X-ray regime (Baskill et al. 2005). The X-ray outburst spectrum has shown hints of the presence of more than one thermal component: a soft (5–30 eV) and a hard one (10 keV). The former originates from the boundary layer although the flux is too low compared to boundary layer models. This may be attributed to a density gradient in the optically thick boundary layer so that a hot optically thin layer is still present and emits hard X-rays (Done & Osborne 1997), though colder and weak (Popham 1999). The hard X-ray flux instead recovers the quiescent level at the end of the optical outburst. Only a few systems had intense multi-wavelength monitoring covering multiple outbursts (Collins & Wheatley 2010). Some DNe show an optical/X-ray anti-correlated behaviour, where the hard X-ray flux is suppressed during optical outburst. Others, like SS Cyg (Wheatley et al. 2003; McGowan et al. 2004) or the old nova and magnetic CV GK Per (Evans et al. 2009), have also shown increases of hard X-rays during the transition phase from optical quiescence to outburst as well as during the fading of the outbursts. However there are DNe, like U Gem (Güver et al. 2006) or GW Lib (Byckling et al. 2009), that do not show hard X-ray suppression, but X-ray enhancement during optical outburst. SS Cyg also showed similar behaviour during anomalous outbursts. The diversity of behaviours depends on several fundamental parameters (Narayan & Popham 1993), but still far from being understood. A recent proposal that diversity might be related to a critical value of the accretion rate that divides optically thin and optically thick BL regimes (Fertig et al. 2011) needs to be assessed with observations of a large sample of DNe.

Besides the onset of DNOs and QPOs during outburst, DNe have shown another striking similarity to transient LMXBs, further suggesting universal relations among accreting WDs, NS and BHs. The surprising detection of a transient radio emission in SS Cyg itself during outburst (Koerding et al. 2008) was interpreted as the onset of a radio jet. DNe were believed to be unable to power jets. The inferred jet power and the relation to the outburst cycle were found analogous to those seen in XRBs, thus suggesting that the disc-jet coupling mechanism is ubiquitous. The additional detection of radio emission from V3885 Sgr, a CV accreting at a high rate comparable to SS Cyg (Koerding et al. 2011), seems to support that radio jets also occur in both steady and transient CVs. It has been recently shown that nova-like CVs can have optically thin inefficient accretion disc boundary layer (ADAF-like) (Balman et al. 2014), allowing the possibility to retain energy to power jets. If this were the case, a radio/X-ray correlation, similar to that observed in NS and BH binaries (Coriat et al. 2011), should be present in these systems.

So far, a handful of DNe have been observed through multiple outbursts and it is crucial to observe systems spanning a range of orbital periods with adequate multi-outburst coverage. Bright outbursts such as those displayed by SS Cyg or GK Per will be easily detected by the WFM and then followed-up by the LAD (Fig. 3, left panel), allowing spectral parameters to be determined at accuracy better than 4%. Fainter sources down to the flux levels as that of GW Lib ($\sim 2 \times 10^{-11}$ erg cm$^{-2}$ s$^{-1}$) can be monitored by the LAD from their early phases onwards through ToO alerts from optical ground-based telescopes. It will be possible to recover the hard X-ray suppression and re-appearance to understand the diversity of behaviours. A crucial determination will be the WD mass through the study of the hard X-ray tails. Short orbital period DNe ($\lesssim 2$ hr) have long (several yrs) duty cycles and long (months) outburst durations, while long ($\gtrsim 3$ hr) orbital period DNe display shorter durations (~weeks) and duty cycles (months) of outbursts. The monitoring of a sample of ~10 DNe during *LOFT* lifetime, with multiple outburst coverage for those with short duty cycles, will allow to characterise not only the spectral evolution of the outbursts, but also the non-periodic variations (see Sect. 3) with *unprecedented details* in the hard domain. Furthermore, proper ToO alerts with radio facilities will be crucial to explore whether the X-ray and radio luminosities are related, thus proving the suspected disc-jet connection *universal* among LMXBs and CVs.

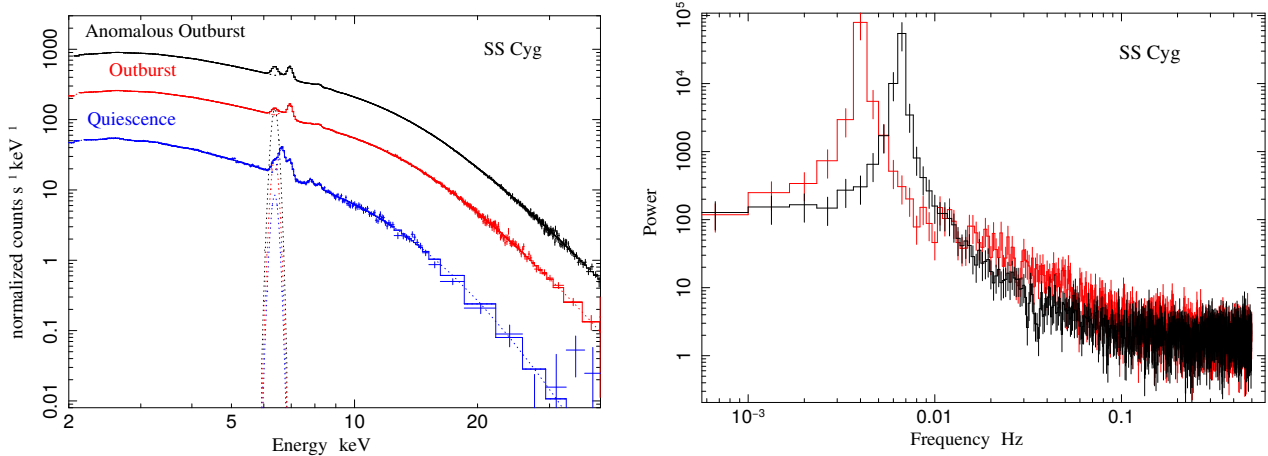

Figure 3: *Left:* From top to bottom: Simulated 10 ks LAD spectra of SS Cyg during anomalous and normal outbursts and in quiescence ($F_{3-20\,keV} = 7, 2, 0.3 \times 10^{-10}$ erg cm$^{-2}$ s$^{-1}$, respectively). Spectra are based on McGowan et al. 2004 that include a thermal component (*mekal*) and a gaussian at 6.4 keV (EW = 82 eV). The temperature varies from 26 keV to 20 keV in outbursts decreasing to 8 keV in quiescence, recovered at $\lesssim 4\%$ level. *Right:* Simulated 6 ks power spectra of SS Cyg during two phases of X-ray turn-on during a normal outburst ($F_{3-20\,keV} = 2, 1.5 \times 10^{-10}$ erg cm$^{-2}$ s$^{-1}$, red and black lines respectively). A broken power law with break frequency $2 \times 10^{-2}$ Hz is assumed. The QPOs have periods of 155 s (black line) and 245 s (red line) and rms amplitudes of 40% and 30%, respectively (after Wheatley et al. 2003).

## 6 Context in post-2020 time frame

Most of the science goals that can be achieved by *LOFT* will greatly benefit from foreseen synergies among facilities in the post-2020 time frame in different energy domains.

- **Population of galactic hard X-ray sources**: Recent studies of GRXE (Revnivtsev et al. 2009; Revnivtsev et al. 2011; Yuasa et al. 2012; Warwick 2014) and galactic bulge (Hong 2012) have revealed the contribution of discrete faint sources that above 2–3 keV are most identified as mCVs. This implies that the galactic X-ray source composition should take into account a large population of faint hard CVs. The *Swift*/BAT and *INTEGRAL*/IBIS hard X-ray surveys have indeed revealed that ~ 7–8% of detected sources (~20% of galactic sources) are accreting WD binaries, most of the magnetic type. Surprisingly, a number of hard X-ray emitting non-magnetic CVs (most symbiotics) were also identified. These could be systems harbouring massive WDs. They are likely highly absorbed, making *LOFT* an ideal facility to study them. About $10^4$ new candidates will be discovered in future X-ray surveys with, e.g., *eROSITA* (Schwope 2012) and carefully selected hard samples at flux levels reachable by *LOFT* will need follow-up X-ray observations. At that time, also precise distances to be provided by the *Gaia* satellite will be available for thousands of these binaries and will be of crucial importance to determine X-ray luminosities and thus understanding the role of accreting WD binaries (both magnetic and non-magnetic) in galactic populations of X-ray sources.

- **Optical/IR coordination**: Optical/nIR ground-based telescopes will be important to provide multi-wavelength coverage. This is especially crucial for CV outbursts (novae and DNe) where optical coordination is required to trigger *LOFT* LAD pointings. The Catalina Real-time Transient Survey (CRTS) has already uncovered ≳1000 new WD binaries and the Palomar Transient Factory (PTF) will come up with large numbers as well. At the time of *LOFT* operations the ultimate variability survey for target selection and characterization will be the Large Synoptic Survey Telescope (LSST). Furthermore simultaneous follow-ups with *LOFT* will be possible with already existing and with planned robotic observatories such as the Stellar Robotic (STELLA) telescope in the northen hemisphere, the Monitoring Network of Telescopes (MONET) in the southern hemisphere and the Las Cumbres Observatory Global Telescope Network (LCOGT). Furthermore ESA's PLATO mission will be also operational during *LOFT*'s life-time and will provide detailed optical light curves for extended periods of time. This will open the possibility to have simultaneous X-ray and optical monitoring of CVs in outbursts and in quiescence with an unprecedented long-term coverage. The expected several thousands of CVs to be detected in the forthcoming decade will require massive ground-based spectroscopic follow-ups, that will be provided by large area facilities such as the 4 m Multi-Object Spectroscopic Telescope (4MOST) at ESO or the Large sky Area Multi-Object Fibre Spectroscopic Telescope (LAMOST) or the Wide-field multi-object spectrograph (WEAVE) in the northern hemisphere. These facilities will allow proper selection of target samples to be investigated by *LOFT*. Time-resolved optical/nIR spectroscopic and polarimetric coverage as provided by the ESO VLT and by the E-ELT in mid-2020 will be also required.

- **Radio coordination**: The onset of transient radio-jet recently claimed in high accretion rate non magnetic CVs is a new challenging aspect that deserves radio coverage to enforce the proposal of a possible universal connection between disc accretion and jet formation in XRBs. In this respect, if and how the X-ray and radio luminosities correlate has to be investigated. Furthermore, the detailed analysis of the evolving radio emission from a recent Fermi detected nova has demonstrated the crucial role of multi-band coordination to understand the mechanisms of unexpected very high energy emission in nova explosions. Radio facilities like the Square Kilometre Array (SKA) will be operational in the *LOFT* timeframe and will be of unique importance in this context.